\documentclass[12pt]{article}
\usepackage{epsfig}
\begin{document}

\title{\LARGE{\bf Quark-Hadron picture of the Deuteron Photodisintegration} }
\date{ }
\maketitle
\vspace{0.001cm}
\centerline{\large M.Mirazita~\footnote{Invited talk at the XL International Winter Meeting on Nuclear Physics, 21-26 January 2002}}
\centerline{\large \it{I.N.F.N. - Laboratori Nazionali di Frascati}}

\vspace{0.5cm}

\begin{center}
{\bf Abstract}
\end{center}
{\it {The high energy two-body photodisintegration of the deuteron is an ideal reaction for the identification of quark effects in the nuclei. In particular, the study of this reaction in the few GeV region can clarify the transition from the nucleonic degrees of freedom to the QCD picture of hadrons. 
The CLAS large angle spectrometer of Hall B at TJNAF allowed for the first time to measure the differential cross section for photon energy between 0.5 and 3 GeV, over all proton angles between 10$^\circ$ and 140$^\circ$ in the laboratory frame. 
Preliminary results from the analysis of 30$\%$ of the total statistic accumulated is shown. These results are in good agreement with the available data, and are well described by the recent calculation of the deuteron photodisintegration cross section derived in the framework of the Quark Gluon String Model.} }

\vspace{0.5cm}

\section{Introduction}
One of the primary goal of nuclear physics is the study of the interplay between hadronic and partonic degrees of freedom, and of the effectiveness of the traditional nucleon-nucleon theories or QCD inspired model in describing the data.
For this purpose, the deuteron reactions with real and virtual photons are particularly well suited, because the deuteron is the simplest nucleus, and photon interaction with quarks is well known. 
\\
Then, a broad program of studies of this kind of reaction has been started at Thomas Jefferson National Accelerator Facility (TJNAF) \cite{holt}. 
It includes measurements of: differential cross section and recoil proton polarization in the two-body deuteron photodisintegration, meson photoproduction cross sections, electromagnetic form factors in the $ed$ elastic scattering with both unpolarized and polarized electron beams.
\\
Exclusive reactions at low energies (up to around 0.5 GeV) are usually correctly described in terms of hadronic degrees of freedom and meson exchange currents (MEC). 
At higher energies, the QCD structure of hadrons becomes dominant, and data are reproduced using perturbative QCD calculations.
The intermediate energy region is important in order to clarify the transition from the hadronic to the partonic description of hadrons and nuclei. However, in this region perturbative QCD calculation are not applicable, thus non-perturbative methods must be used.
Moreover, in the photonuclear reactions the highest momentum transfer $p_T$ can be obtained, because the absorbed photon delivers all of its energy to the interacting quarks. Then, one can expects that QCD effects become to manifest at lower energies in deuteron photodisintegration, compared for example with $ed$ scattering.

\section{Available data and theories}

\subsection{The differential cross section}

\begin{figure}[ht]
\begin{center}
\epsfig{file=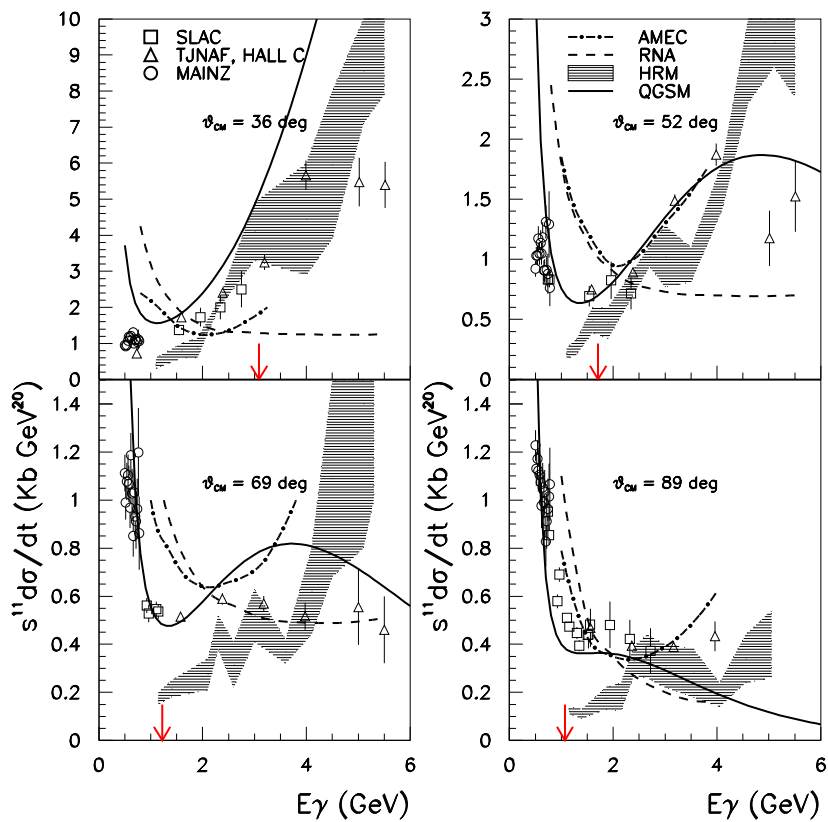,width=12cm}
\caption{\it {Deuteron photodisintegration cross section multiplied by $s^{11}$. Experimental data are from Mainz \cite{mainz}, SLAC \cite{slac} and Hall C of TJNAF \cite{tjnafexp}. The arrows indicate where the onset of CCR scaling is expected \cite{ccr}. For the theoretical curves, see the text.}}
\label{fig1}
\end{center}
\end{figure}

One obvious signature of QCD effects in the deuteron photodisintegration is a scaling behaviour of the cross section.
The constituent counting rule (CCR) \cite{ccr} predicts that, for sufficiently high energy and fixed angle, the differential cross section of any binary reaction must scale with the square of the total energy $s$:
\[\frac{d\sigma}{dt} = \frac{1}{s^{N_F - 2}} f(\theta_{CM})\]
where $N_F$ is the minimum number of microscopic fields involved in the reaction. In our case, $N_F = 13$, thus $d\sigma/dt \propto s^{-11}$.
\\
The differential cross section of the deuteron photodisintegration for photon energy above 1 GeV has been measured only at few angles \cite{mainz}, \cite{slac}, \cite{tjnafexp}. The available experimental data are shown in fig. \ref{fig1} for four proton angles in the center of mass frame. 
In order to evidentiate the possible CCR scaling, the cross section is here already multiplied by $s^{11}$, and the arrows in the plots indicate the expected threshold for the onset of the scaling, $p_T > $1 GeV \cite{ccr}.
However, the experimental data show scaling only at large proton angle, $\theta_{cm} = 69^{\circ}$ and $89^{\circ}$, while for more forward angles the situation is less clear. 
\\
More sophisticated models have been developed for the description of the data. In the Reduced Nuclear Amplitude (RNA) model \cite{rna} the binding of the quarks inside the nucleons and the deuteron is taken into account with empirical form factors and the elementary cross section is computed assuming CCR scaling.
Though this approach successfully reproduce the $ed$ elastic scattering even at energies well below the expected onset of the CCR scaling \cite{rnael}, it is able to describe the $\gamma d \rightarrow p n$ cross section with an appropriate normalization factor only for $\theta_{cm} = 69^{\circ}$ and $E_{\gamma} >$ 2 GeV.
\\
In the Hard quark Rescattering Model (HRM) \cite{hrm}, it is supposed that the main contribution to the interaction comes from the absorption of the photon by a quark of a nucleon, which subsequently interacts with a quark of the other nucleon with high momentum transfer, producing the final state with two nucleons with high relative momentum. 
The limitation for the applicability of this model are $E_{\gamma} >$ 2.5 GeV and t $>$ 2 GeV$^2$, but, under particular assumptions for the short distance $pn$ interaction, it can be extended beyond its limits. However, the agreement with the data is poor, especially at higher energies, and the uncertainty is large, due to the poor knowledge of the $pn$ amplitude.
\\
A non perturbative QCD calculation has been done in the framework of the Quark Gluon String Model (QGSM) \cite{gris}. 
It assumes that the scattering amplitude at high energy is dominated by the exchange of three quarks in the t-channel, and the duality property allows its extension at lower energies.
The intermediate three quarks state con be identified with the nucleon Regge trajectory, which is derived from recent QCD calculations. 
More details and comments on this model are given in Section 4.
\\
Traditional models based on hadronic degrees of freedom are expected to fail to reproduce data at $E_{\gamma} >$ 1 GeV. 
However, this approach has been extended in the few GeV region in the Asymptotic Meson Exchange Current (AMEC) model \cite{amec}, using form factors to describe the $dNN$ interaction vertex and an overall normalization factor fixed by fitting the experimental data at 1 GeV.
The results reproduce the energy dependence of the cross section only for $\theta_{cm} = 89^{\circ}$. 
In any case, it is worth noticing that also this non QCD-based model is able to provide a scaling law for the cross section, with an exponent depending on the scattering angle.

\subsection{The proton polarization}
\begin{figure}[ht]
\begin{center}
\epsfig{file=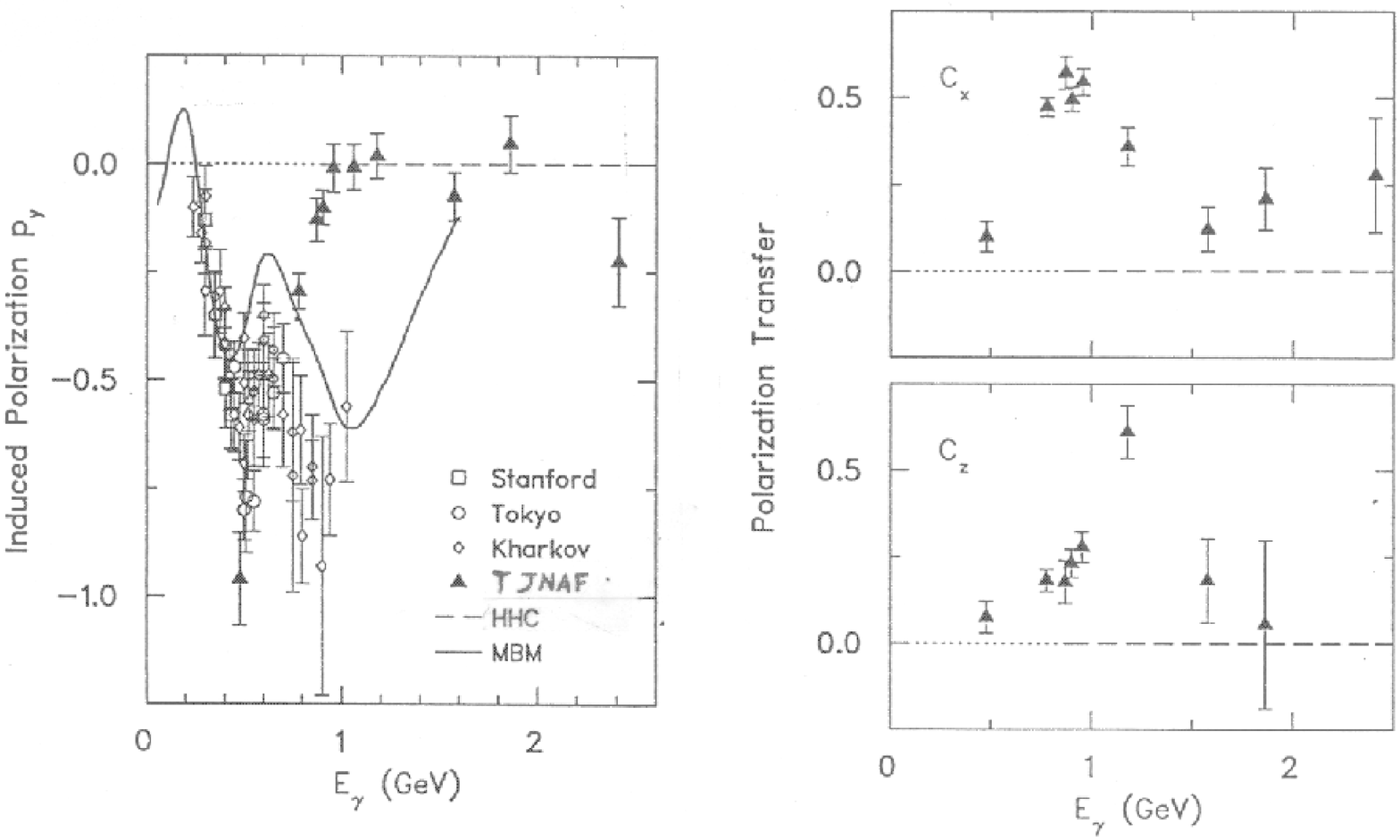,width=12cm}
\caption{\it {Induced proton polarization $p_y$ and polarization transfer $C_x$ and $C_z$ in the deuteron photodisintegration at $\theta_{cm} = 90^{\circ}$ (\cite{pol} and references therein). The y-axis is perpendicular to the scattering plane, the z-axis is parallel to the scattered proton momentum and the x-axis is perpendicular to y and z.}}
\label{fig5}
\end{center}
\end{figure}

Polarization observables provide informations complementary to the differential cross section. 
At high energies, a typical signature of pQCD effects is the Hadron Helicity Conservation (HHC) \cite{hhc}, that arises from the fact that photon-quark and quark-gluon interactions conserve the chirality, if quark masses can be neglected. Then, the total quark helicity is conserved.
\\
So far, no exclusive reactions have been shown to satisfy HHC, but the number of studies is limited.
The most detailed investigations have been done in $pp$ elastic scattering \cite{hhcprot}, which is approximately consistent with CCR, but not with HHC. 
This has been attributed to long distance phenomena in which there are three independent scattering of quarks in the beam proton from quarks in the target proton \cite{landmecprot}. 
This mechanism is strongly suppressed in photoreactions \cite{landmecgam}, because the incoming photon can only interact with a single quark in the target, thus one can expect HHC in photoreactions.
\\
The proton polarization has been measured in the $d(\vec{\gamma}, \vec{p}) n$ reaction only for the proton angle $\theta_{cm} = 90^{\circ}$ \cite{pol}, and the results are shown in fig. \ref{fig5}.
For the normal component of the polarization $p_y$, data are consistent with the Meson-Baryon exchange Model (MBM) up to 0.5 GeV. 
Above 1 GeV, the recent data taken at Hall A of TJNAF show that the induced polarization is small, consistent with HHC. However, the transverse $C_x$ and longitudinal $C_z$ in-plane polarization are non-zero, and indicate that helicity is not conserved at least up to 2 GeV.
\\
We thus can conclude that experimental data at $\theta_{cm} = 90^{\circ}$ on cross section and polarization give results that can't be consistently interpreted in a single theoretical picture.
Clearly, more data for other proton angles are needed for both the observables, in order to provide stronger constraints on the theoretical models.

\section{The experiment E93-017 at TJNAF}
\begin{figure}[htp]
\begin{center}
\epsfig{file=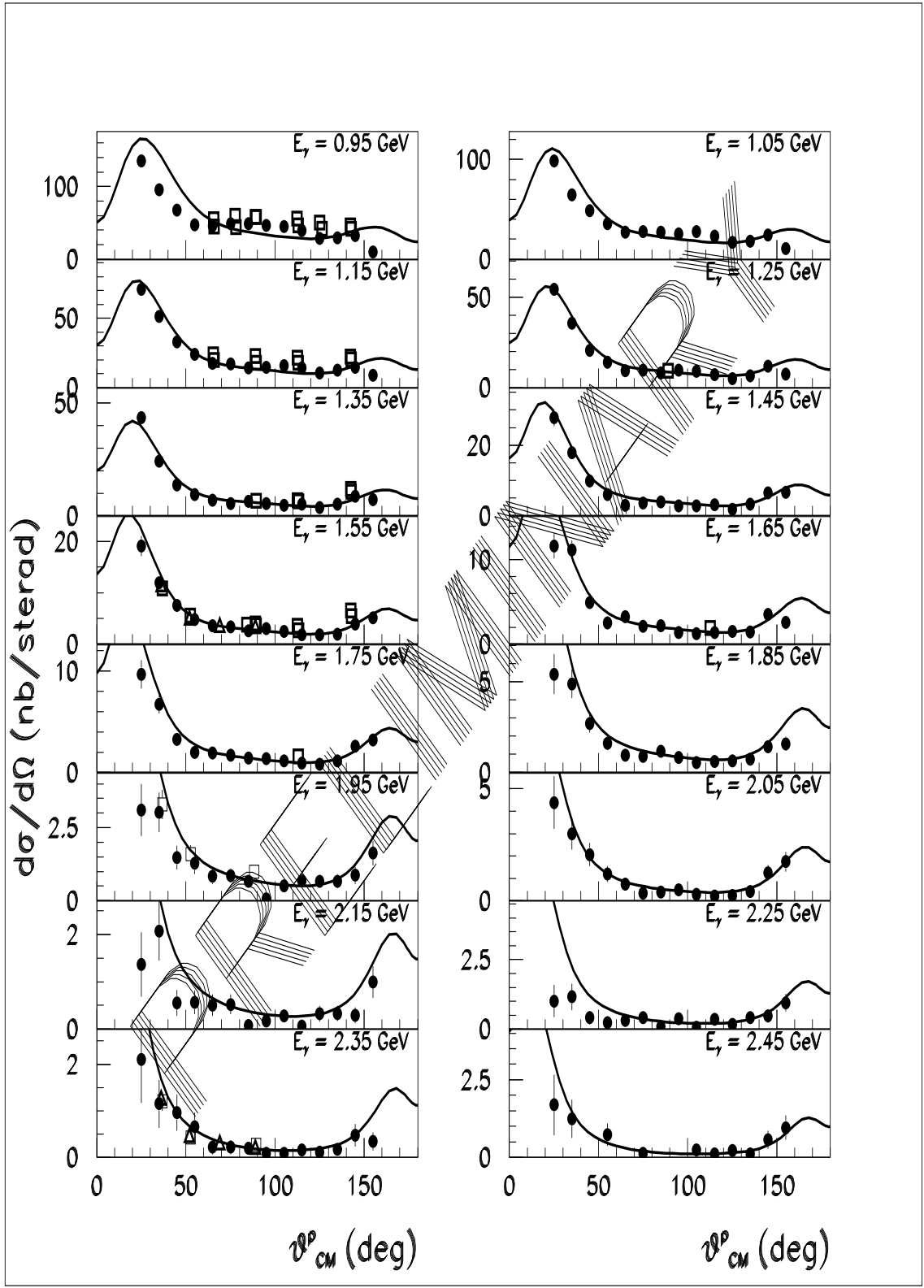,width=15cm,height=21cm}

\caption{\it {Preliminary results of the deuteron photodisintegration differential cross section measured in Hall B of TJNAF (black circles), compared with the published data from SLAC \cite{slac} (open squares) and Hall C of TJNAF \cite{tjnafexp} (open triangles). The curve is the QGSM calculation \cite{gris}.}}
\label{fig2}
\end{center}
\end{figure}

New data of the deuteron photodisintegration cross section are obtained at TJNAF \cite{prop} with the CEBAF Large Angle Spectrometer (CLAS) of Hall B. A bremsstrahlung photon beam is produced by the continuous electron beam hitting a thin radiator, and the tagging system \cite{tag} tags monochromatic photons with energy between 0.20 and 0.95 of the electron beam and resolution of about 0.1$\%$. 
The hadrons are detected with CLAS \cite{clas}, a nearly 4$\pi$ spectrometer based on a toroidal magnetic field generated by 6 superconducting coils.
The field was set to bend positively charged particles away from the beam line. 
The 6 coils define 6 independent modules. Each module is implemented with 3 regions of drift chambers \cite{dc} for the tracking of charged particles, and time-of-flight (TOF) scintillators \cite{tof} for charged hadron identification. 
Other detectors of CLAS (not used for the present analysis) are two electromagnetic calorimeters, covering angles up to 70 degrees, for the detection of neutral particles, and Cerenkov counters for $e/\pi$ separation.
The resolution of the proton momentum is of the order of a few percent, while the proton acceptance is close to 90$\%$ in the fiducial region of the detector.
\\
Data were taken with photon energy ranging from 0.5 to 3 GeV. 
The trigger for the data acquisition was the coincidence between signals in the tagger and in the CLAS detector (TOF). The total number of recorded triggers is about 2.5 billions.
\\
Real photodisintegration events are selected requiring the identification of a photon in the tagger and a proton in CLAS, and then applying a missing mass cut to the reaction $\gamma d \rightarrow p X$.
The CLAS acceptance has been evaluated by Monte Carlo simulations of the photodisintegration reaction, and the background contribution has been computed by a fit to the experimental missing mass distributions.
\\
In fig. \ref{fig2} the differential cross section $d\sigma / d\Omega$ is reported as a function of the proton angle in the CM frame, for fixed photon energy above 0.9 GeV. 
The present preliminary results are obtained from the analysis of about $30\%$ of the accumulated statistic, corresponding to about 530 millions of triggers.
Moreover, an additional cut $\theta_p^{CM} > 20^{\circ}$ has been applied, that will be removed in the final analysis. 
The agreement with the other available data from SLAC \cite{slac} and Hall C of TJNAF \cite{tjnafexp} is good at all energies and angles. 
Notice that, in spite of the limited data set, the statistical error is lower than $5\%$ for $E_\gamma < 1.5$ GeV. When the whole data set will be analysed, the statistical error will be improved by a factor 3 or more for photon energies up to 2.5 GeV.

\section{The Quark Gluon String Model}
\begin{figure}[htb]
\begin{center}
\epsfig{file=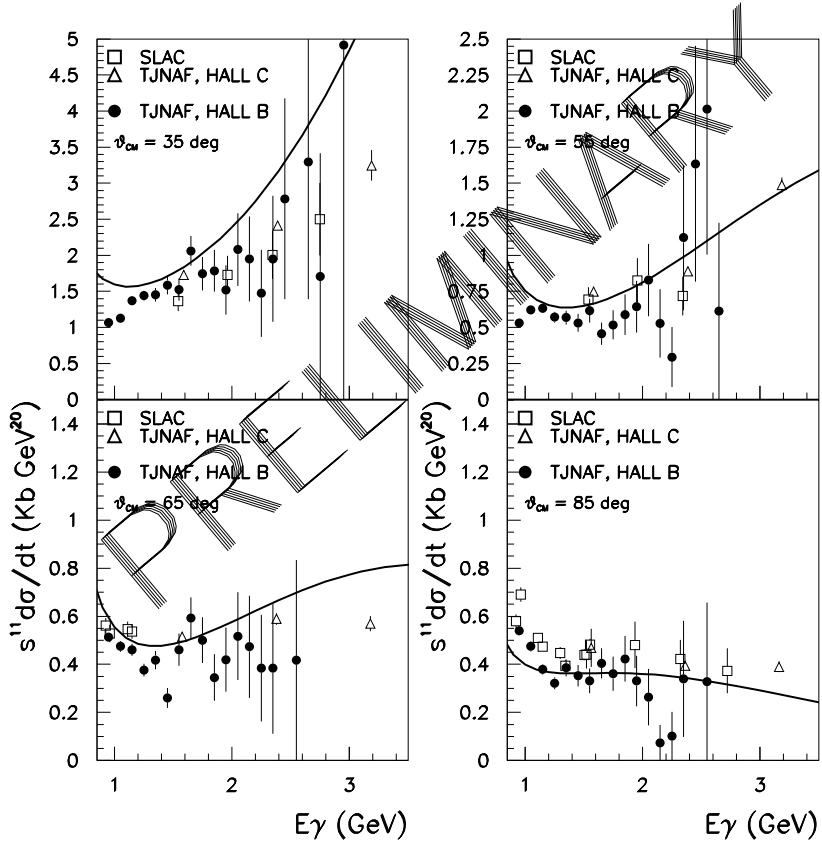,width=12cm}
\caption{\it {Deuteron photodisintegration cross section for four fixed proton angles: preliminary results of Hall B of TJNAF (black circles) and published data from Mainz \cite{mainz}, SLAC \cite{slac} and Hall C of TJNAF \cite{tjnafexp}. The curve is the QGSM calculation \cite{gris}.}}
\label{fig4}
\end{center}
\end{figure}

The Quark Gluon String Model (QGSM) is a non-perturbative approach, which has been extensively used for the description of hadronic reactions at high energies \cite{kaida}. 
Due to duality property of scattering amplitudes, it can also be applied at intermediate energies for reactions without explicit resonances in the direct channel. 
In fact, it well describes the reactions $p p \rightarrow d \pi^+$ and $\bar{p} d \rightarrow M N$, where diagrams with the exchange of three valence quarks in the t-channel are dominant \cite{qgsmadr}.
This approach has also been used for the description of heavy ions collisions at high energy \cite{ions}.
\\
The QGSM is based on a topological expansion in QCD of the scattering amplitudes in power of $1/N$ \cite{1/n}, where $N$ is the number of colors $N_c$ or flavors $N_f$. In the limit $N_f >> 1$ and $N_c/N_f \simeq 1$, the planar graphs give the dominant contribution to the amplitude \cite{vene}. 
In the case of exclusive reactions, such as the deuteron photodisintegration, the actual expansion parameter is $1/N^2 = 1/9$, thus the expansion is expected to work.
\\
In the space-time representation, the reaction $\gamma d \rightarrow p n$ is described by the exchange of three valence quarks in the t-channel with any number of gluons between them. This picture corresponds to the formation and break up of a quark-gluon string in the intermediate state, leading to the factorization of the amplitudes: the probability for the string to produce different hadrons in the final state does not depend on the type of the annihilated quarks, but is only determined by the flavor of the produced quarks.
\\
The intermediate quark-gluon string can be easily identified with the nucleon Regge trajectory \cite{kaida}. Actually, most of the QGSM parameter can be related to the parameters of Regge Theory, as trajectories or residues. In this sense, the QGSM can be considered as a microscopic model for the Regge phenomenology, and can be used for the calculation of different quantities that have been considered before only at a phenomenological level.
\\
In Ref. \cite{gris}, the QGSM has been applied for the description of the deuteron photodisintegration reaction, using QCD motivated non linear nucleon Regge trajectories \cite{brisu}, with full inclusion of spin variables and assuming the dominance of the amplitudes that conserve s-channel helicity. 
The interference between the isoscalar and isovectorial components of the photon has been also taken into account, leading to forward-backward asymmetry in the cross section. 
\\
The result of the calculation is shown in fig. \ref{fig2} as full line, compared with previous published data and with the preliminary results of the new CLAS measurement.
The agreement between data and calculations is very good for all energies above 1 GeV: the QGSM reproduces the angular dependance of the cross section at fixed photon energy and predicts the forward-backward asymmetry showed by the data. 
From these plots, it results also the importance of measuring the cross section at very forward angles, in order to check the QGSM prediction that the cross section decreases for angles smaller than $10^{\circ}-20^{\circ}$.
\\
The QGSM model also predicts correctly the decrease of $d\sigma / dt$ at fixed angle as a function of the photon energy, as shown in fig. \ref{fig4} for four CM proton angles. Also here, the cross section multiplied by the factor $s^{11}$.

\section{Conclusions}
For its simplicity, the deuteron is a very interesting system for the study of the quark structure of nuclear matter. 
In particular, the deuteron photodisintegration $\gamma d \rightarrow p n$ is an ideal reaction, because photon-quark interactions are well known and large momentum transfer can be obtained.
\\
The more interesting result of the published experimental data is that in the deuteron photodisintegration at $90^{\circ}$ the cross section is consistent with the CCR scaling law, while the measured proton polarizations $C_x$ and $C_z$ show that HHC is not conserved at least up 2 GeV.
\\
New measurements of the cross section have been performed in the Hall B of TJNAF, using photons with energy between 0.5 and 3 GeV, covering almost all proton angles in the proton CM frame.
The preliminary results of about of $30\%$ of the total statistic accumulated and photons with energy between 0.9 and 2.5 GeV are in good agreement with the already published data.
\\
The experimental data have been analysed in the framework of the QGSM, which is a non perturbative approach extensively used for the description of binary hadronic reactions and of heavy ions collisions at high energy. 
This model provides a differential cross section that reproduces the data for all proton angles and photon energy above 1 GeV.
QGSM calculations of the proton polarization are underway and will be soon available. 
Two new measurements of polarization observables at high energies and different proton angles are planned at TJNAF, and they will be a further, strong test of the QGSM predictions.

\end{document}